\documentclass[twocolumn, secnumarabic, amssymb, nobibnotes, aps, prd]{revtex4-1}

\usepackage{graphicx}
\usepackage{dcolumn}
\usepackage{bm}

\begin{document}
\title{Effects of $^{3}$He impurities on the mass decoupling of $^{4}$He films}

\author{Kenji Ishibashi$^{a}$}
\author{Jo Hiraide$^{a}$}
\author{Junko Taniguchi$^{a}$}
\author{Tomoki Minoguchi$^{b}$}
\author{Masaru Suzuki$^{a}$}%
\email{suzuki@phys.uec.ac.jp}

\affiliation{%
$^{a}$Department of Engineering Science, University of Electro-Communications,
Chofu, Tokyo 182-8585, Japan. \\
$^{b}$Institute of Physics, University of Tokyo, Meguro-ku, Tokyo 153-8902, Japan
}%

\date{\today}

\begin{abstract}
We carried out quartz crystal microbalance experiments of a 5~MHz AT-cut 
crystal for superfluid $^{4}$He films on Grafoil (exfoliated graphite)
with a small amount of $^{3}$He up to 0.40~atoms/nm$^{2}$. 
We found that the mass decoupling from oscillating substrate is  
considerable sensitive even in a small amount of $^{3}$He doping. 
In a $^{4}$He film of 29.3~atoms/nm$^{2}$, we observed a small drop 
in resonance frequency at $T_{3}$ of $\sim$0.4~K for a small amplitude, 
which is attributed to sticking of $^{3}$He atoms on the $^{4}$He solid 
atomic layer. For a large amplitude, the $^{4}$He solid layer shows 
a reentrant mass decoupling at $T_{R}$ close to $T_{3}$. 
This decoupling can be explained by the suppression of the superfluid 
counterflow due to the adsorption of $^{3}$He atoms on edge dislocations.  
As the $^{4}$He areal density increases, $T_{R}$ shifts to the lower 
temperature, and vanishes around a $^{4}$He film of 39.0~atoms/nm$^{2}$. 
\end{abstract}

\maketitle

\narrowtext
\section{Introduction} 
%
It is well known that the surface of graphite is atomically flat 
and helium film on graphite grows up layer-by-layer to more than 
five-atom thick film in layers.\cite{Greywall_1991,Crowell_1996} 
Because of both the quantum nature of helium and the ideal 
two-dimensional system, helium film on graphite has been attracting 
the attention of many researchers. The adsorbed 
structure,\cite{Greywall_1991,Pierce_1999,Corboz_2008} 
the magnetism\cite{Neumann_2007,Fukuyama_2008} 
and the superfluidity\cite{Crowell_1996,Nyeki_2017} are extensively 
studied experimentally and theoretically. 

Lately the nanofriction of films, or the mass decoupling of films  
from oscillation, has been widely discussed.\cite{Krim_2012} 
Several films on metal substrates 
show a partial mass decoupling.\cite{Dayo_1998}. In addition, it is reported 
that the film takes place the pinning-depinning transition against 
the driving force of oscillating substrates.\cite{Bruschi_2002}

In response to the study on nanofriction, we started to study the mass 
decoupling of helium films on graphite using the quartz crystal 
microbalance (QCM) technique. 
Up to the present, we have reported the following observations above 
two-atom thick films: 
\cite{Hosomi_2007,Hosomi_2008,Hosomi_2009,Hosomi_2007A} \\
(a) When the oscillation amplitude is large enough, the solid layer of $^{4}$He 
films undergoes partial mass decoupling below a certain temperature $T_{S}$. \\ 
(b) This decoupling brings a low-friction metastable state when the overlayer 
is normal fluid. The solid layer after the reduction in amplitude remains 
in the low-friction state with a finite life time. \\
(c) When the overlayer is superfluid, the mass decoupling suddenly vanishes 
at $T_{D}$ below $T_{S}$.\\
(d) For $^{3}$He films, the mass decoupling shows a similar behavior up 
to five-atom-thick films without an abrupt suppression due to superfluid. 

The inhomogeneity of films plays an important role in this decoupling. 
We proposed the following scenario:\cite{Hosomi_2009}
The motion of the edge dislocation in the solid layer is responsible for mass 
transport. The mass decoupling occurs when the edge dislocation overcomes 
the potential barriers of the substrate (Peierls potential). 
This explains the external force threshold for mass decoupling and 
the low-friction state being metastable. In addition, the sudden vanishment
below $T_{D}$ can be explained by the cancellation of mass transport due to 
the superfluid counterflow of the overlayer.

The mass decoupling of helium films has shown various interesting behaviors. 
In the present experiments, we confine ourselves to $^{3}$He impurity effects 
of $^{4}$He films on Grafoil (exfoliated graphite) when the overlayer is superfluid. 
%
In this Paper we report a systematic study on the mass decoupling using a MHz range 
AT-cut crystal. 
After a brief explanation on the experimental setup in II, we show In III.1 
the $^{3}$He areal density dependence for a four-atom thick film 
for various oscillation amplitudes. By adding a small amount of $^{3}$He, it was 
found that the mass decoupling appears again at a certain temperature $T_{R}$ 
below $T_{D}$ for a large amplitude. 
In III.2, we show the $^{4}$He areal density dependence for a fixed amount 
of $^{3}$He. $T_{R}$ decreases with increasing the $^{4}$He areal density, and 
disappears above a certain $^{4}$He areal density. 
In addition to these observations, we discuss a possible mechanism of the 
reentrant mass decoupling at $T_{R}$. 

\section{Experimental Setup} 
We used the QCM technique with an AT-cut crystal to measure the mass 
decoupling. In the QCM technique, the coupled mass to the oscillating substrate 
is obtained from the change in the resonance frequency $\Delta f$ as 
\begin{equation}
	\frac{\Delta f}{f} = - \frac{m}{M}
\end{equation}
where $m$ is the coupled mass of film, $M$ is the oscillating mass of the 
crystal, and $f$ is the resonance frequency. When the film is decoupled from 
the oscillation, the coupled mass decreases and the resonance frequency increases. 

In the present experiments, the resonator is a 5.0 MHz AT-cut crystal. 
The crystal was commercially available, and no special treatment was applied 
to the Ag electrode. At first, Grafoil was baked 
in a vacuum at 900$^{\circ}$C 
for 3~h, and a 300-nm-thick film of Ag was deposited onto it. The crystal
and Ag-plated Grafoil were pressed together and were heated in a vacuum 
at 350$^{\circ}$C for 2~h. Then, Grafoil was bonded on both sides of the Ag 
electrode. After bonding, the excess amount of Grafoil was removed 
to increase the $Q$ value of the crystal. To keep good thermal contact, 
the crystal was fixed to the metal holder with electrically conductive adhesive. 
After these processes, the $Q$ value remained better than $10^{4}$, and 
the areal density of Grafoil was 7.30~g/m$^{2}$. After being heated 
in $2\times10^{-6}$~Pa at 130$^{\circ}$C for~5 h, the crystal was mounted 
in the sample cell. In the present experiments, the mass loading of $^{4}$He 
is 3.8~Hz$\cdot$atoms$^{-1}$$\cdot$nm$^{2}$.

The resonance frequency was measured using a transmission circuit. In the 
circuit, the crystal was placed in series with a coaxial line 
connecting a 50~$\Omega$ cw signal generator and a RF lock-in amplifier. 
The frequency of the signal generator was then controlled in order to keep 
the inphase output zero, and was locked to the resonance frequency. The 
quadrature output at this frequency is the resonance amplitude.

In the present experiments, the $^{3}$He areal density is at most up 
to 0.4~atoms/nm$^{2}$, which corresponds to 5\% of the areal density of 
$^{4}$He one-atomic layer.

\section{Results and discussion} 
\subsection{$^{3}$He areal density dependence}

\begin{figure}
\centerline{\includegraphics[width=2.8in]{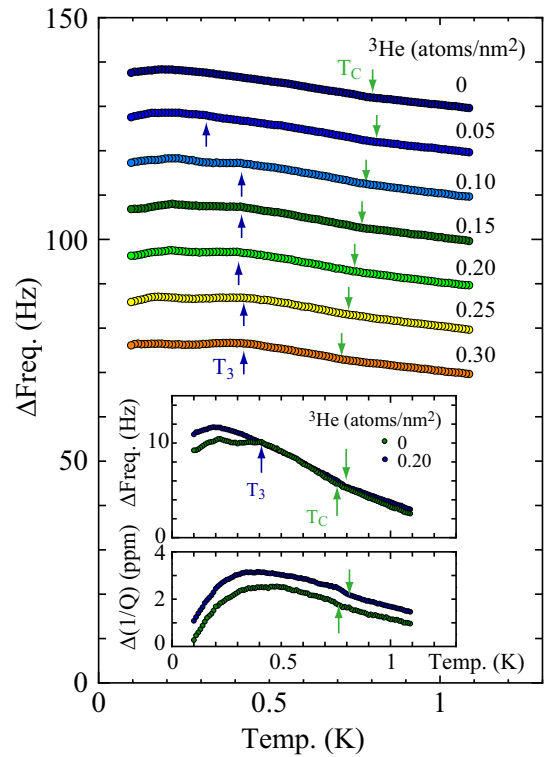}}
\caption{\label{fig:fig1}
	Variations in the resonance frequency of a $^{4}$He film of 29.3~atoms/nm$^{2}$
	at an amplitude of 0.018~nm for various $^{3}$He areal densities. 
	The data are shifted vertically. 
	Inset: Comparison in the resonance frequency and $Q$-value between 
	pure $^{4}$He film and $^{4}$He film with $^{3}$He of 0.20~atoms/nm$^{2}$. 
	(Run~A)} 
\end{figure}

\begin{figure}
\centerline{\includegraphics[width=2.8in]{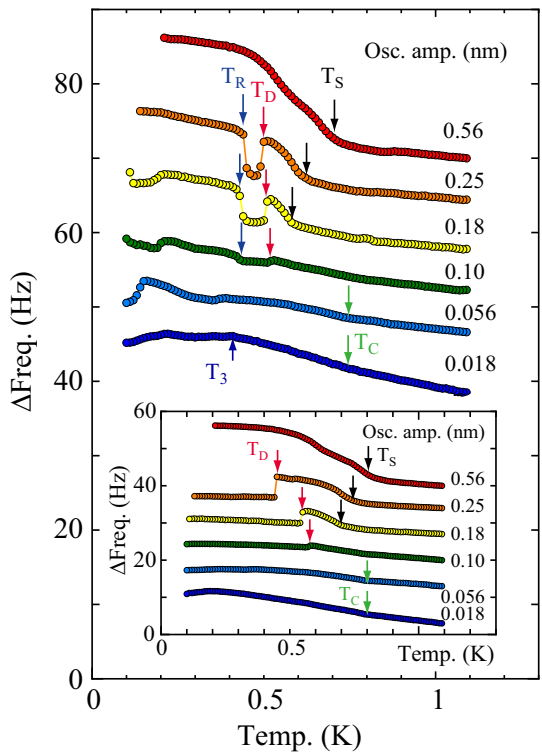}}
\caption{\label{fig:fig2} 
	Variations in the resonance frequency of a $^{4}$He film of 29.3~atoms/nm$^{2}$ 
	with $^{3}$He of 0.20~atoms/nm$^{2}$ for various oscillation amplitudes. 
	The data are shifted vertically. Inset: Variations in the resonance 
	frequency of a pure $^{4}$He film of 29.3~atoms/nm$^{2}$. (Run~A)} 
\end{figure}

We carried out temperature sweep experiments of a four-atom thick $^{4}$He 
film for various oscillation amplitudes by changing the $^{3}$He areal 
density (Run~A). 

Figure~1 shows the variation in resonance frequency for $^{4}$He of 
29.3~atoms/nm$^{2}$ with several $^{3}$He areal densities. 
The overlayer of these films undergoes superfluid at low temperatures. 
All data were taken during cooling with the oscillation
amplitude being fixed at 0.018~nm. In this amplitude, the superfluid onset 
of a pure $^{4}$He film was clearly observed at $T_{C}$ of 0.80~K, 
although it is hardly seen in the scale of Fig.~1. 
As the $^{3}$He areal density increases, $T_{C}$ decreases gradually. 
For a $^{4}$He film with $^{3}$He of 0.30~atoms/nm$^{2}$, $T_{C}$ is 
shifted down to 0.75~K. 
By adding $^{3}$He, it was found that a small additional drop in resonance 
frequency appears at $T_{3}$ below $T_{C}$. As the $^{3}$He areal density 
increases, this drop becomes clear. However, $T_{3}$ does not depend strongly 
on the $^{3}$He areal density above 0.1~atoms/nm$^{2}$. 

In the inset, we compare the variation in resonance frequency and $Q$-value 
between the pure $^{4}$He film and the $^{4}$He film with $^{3}$He of 
0.20~atoms/nm$^{2}$. For the pure $^{4}$He film, the superfluid onset is 
observed at $T_{C}$ of 0.80~K, accompanied with a small increase in $\Delta (1/Q)$. 
When $^{3}$He is added by 0.20~atoms/nm$^{2}$, $T_{C}$ is slightly shifted 
down to 0.76~K. 
The resonance frequency is deviated downwards at $T_{3}$ of 0.41~K from 
the extrapolated curve from high temperatures. The difference from this 
extrapolated curve increases gradually down to the lowest attainable 
temperature $\sim$0.1~K, and becomes $\sim$1.5~Hz. 
On the other hand, the anomaly in 
$\Delta (1/Q)$ was not observed at $T_{3}$ within the present accuracy. 

It is natural that the drop below $T_{3}$ is connected to the addition 
of $^{3}$He. The mass loading of $^{3}$He is estimated to be 
2.9~Hz$\cdot$atoms$^{-1}$$\cdot$nm$^{2}$ from that of $^{4}$He. 
The drop of $\sim$1.5~Hz at low temperature corresponds to 
$\sim$0.5~atoms/nm$^{2}$ for $^{3}$He. This value is about the double of 
$^{3}$He dopant. 
Thus, it is concluded that the drop below $T_{3}$ is caused not only 
by the sticking of $^{3}$He atoms on the $^{4}$He solid layer, but also 
by preventing $^{4}$He atoms from decoupling. 
Furthermore, it was found that $T_{3}$ does not 
depend strongly on the $^{3}$He areal density above 0.1~atoms/nm$^{2}$, which 
means that a number of adsorption sites for $^{3}$He atoms is on the order 
of 0.1~nm$^{-2}$.

The possible candidate of the adsorption site on the $^{4}$He solid  
layer is the edge dislocation core. Because of the adsorption potential 
of graphite, the first solid atomic layer is about 20\% denser than 
the second one,\cite{Greywall_1991} 
Due to the density difference between the solid layers, it is naturally 
assumed that the top solid atomic layer consists of commensurate domains 
separated by domain walls to the first solid atomic layer. 
Since domain walls have the same motif as edge dislocations,\cite{Hosomi_2009} 
we here call them edge dislocations. 
The local areal density of the top solid atomic layer becomes 
small at the edge dislocation. From the difference in the zero-point energy, 
it is thought that $^{3}$He atom is adsorbed on the edge dislocation core from 
the liquid overlayer. Here, it should be noted that the thickness of the 
liquid overlayer is at most one atomic layer and that $^{3}$He atoms 
may not be bounded on the free surface in contrast to bulk $^{4}$He.\cite{Edwards_1978}
In fact, it is revealed that $^{3}$He atoms are trapped on the dislocation 
core with the adsorption potential of 0.7~K in the case of $^{3}$He-$^{4}$He 
solids.\cite{Souris_2015} 

Figure~2 shows that the amplitude dependence for a $^{4}$He film with $^{3}$He 
of 0.20~atoms/nm$^{2}$. All data were taken during cooling. As shown in Fig.~1, 
for the amplitude of 0.018~nm, the superfluid onset and the drop in frequency 
are observed at $T_{C}$ of 0.76~K and $T_{3}$ of 0.41~K, respectively. 
As the amplitude increases, the increase in resonance frequency due to the 
superfluid onset is smeared out. In contrast, for the amplitudes of 0.18, 
0.25 and 0.56~nm, the resonance frequency increases clearly at $T_{S}$, and this 
increase is terminated abruptly at $T_{D}$. 

As shown in the inset, these behaviors are also observed for a pure $^{4}$He 
film, which is attributed to the decoupling and sticking of the $^{4}$He 
solid layer. 
By adding a small amount of $^{3}$He, a new phenomenon appears. 
Below $T_{D}$, the resonance frequency rises up at a certain temperature $T_{R}$, 
which means that the $^{4}$He solid layer undergoes decoupling again. 
It was found that the reentrant mass decoupling temperature $T_{R}$ is 
close to $T_{3}$, i.e., the temperature where $^{3}$He atoms are trapped 
at the adsorption site on the $^{4}$He solid layer. 
For the amplitude of 0.56~nm, $T_{D}$ and $T_{R}$ disappear and the decoupling 
of the $^{4}$He solid layer remains at low temperatures. 

\begin{figure}[t]
\centerline{\includegraphics[width=2.8in]{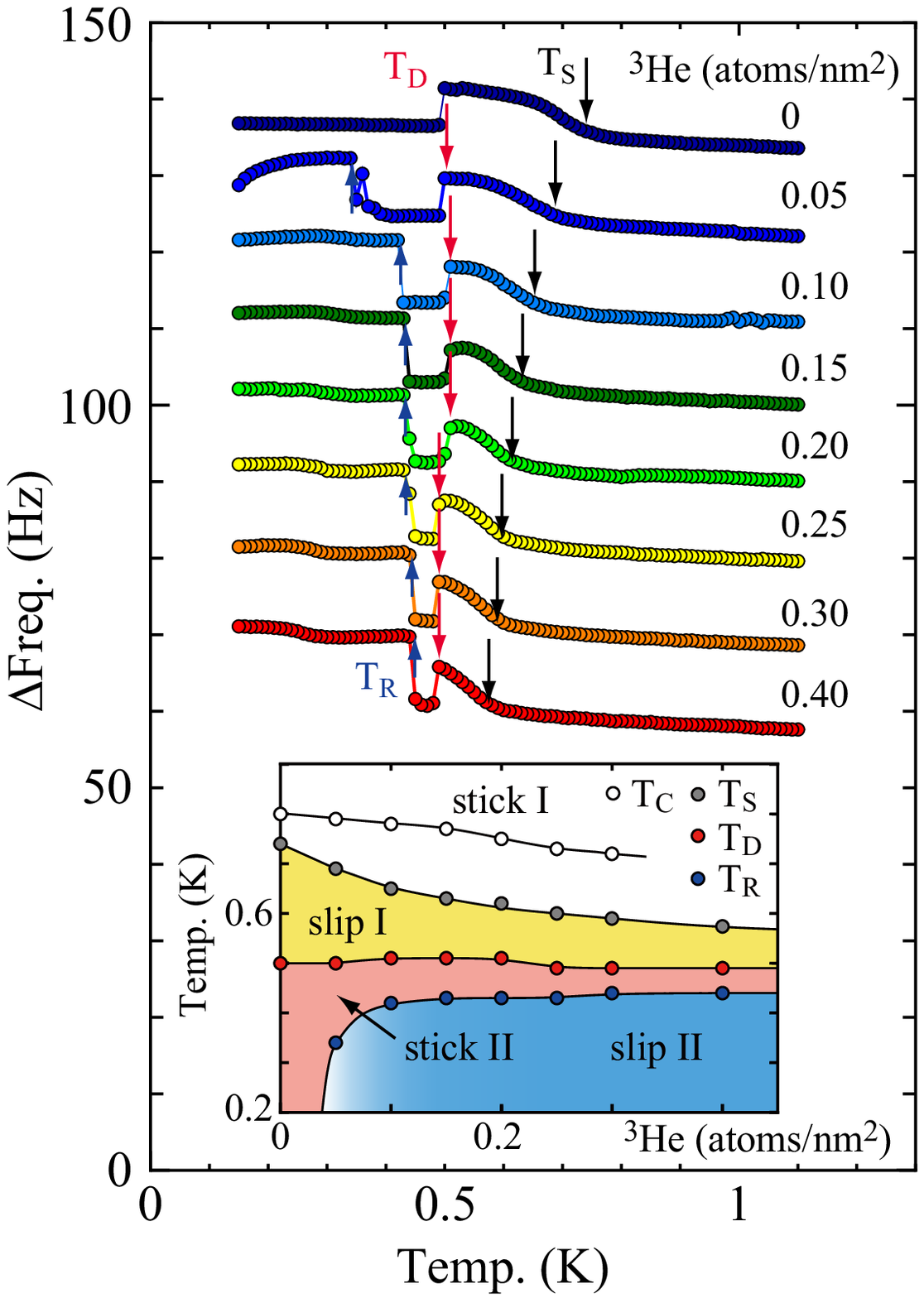}}
\caption{\label{fig:fig3} 
	Variations in the resonance frequency of a $^{4}$He film of 29.3~atoms/nm$^{2}$
	at the amplitude of 0.25~nm for various $^{3}$He areal densities. 
	The data are shifted vertically.
	Inset: Phase diagram of the decoupling and sticking behaviors. 
	$T_{S}$, $T_{D}$ and $T_{R}$ are from Fig.~3, while $T_{C}$ from Fig.~1. 
	(Run~A)} 
\centerline{\includegraphics[width=2.8in]{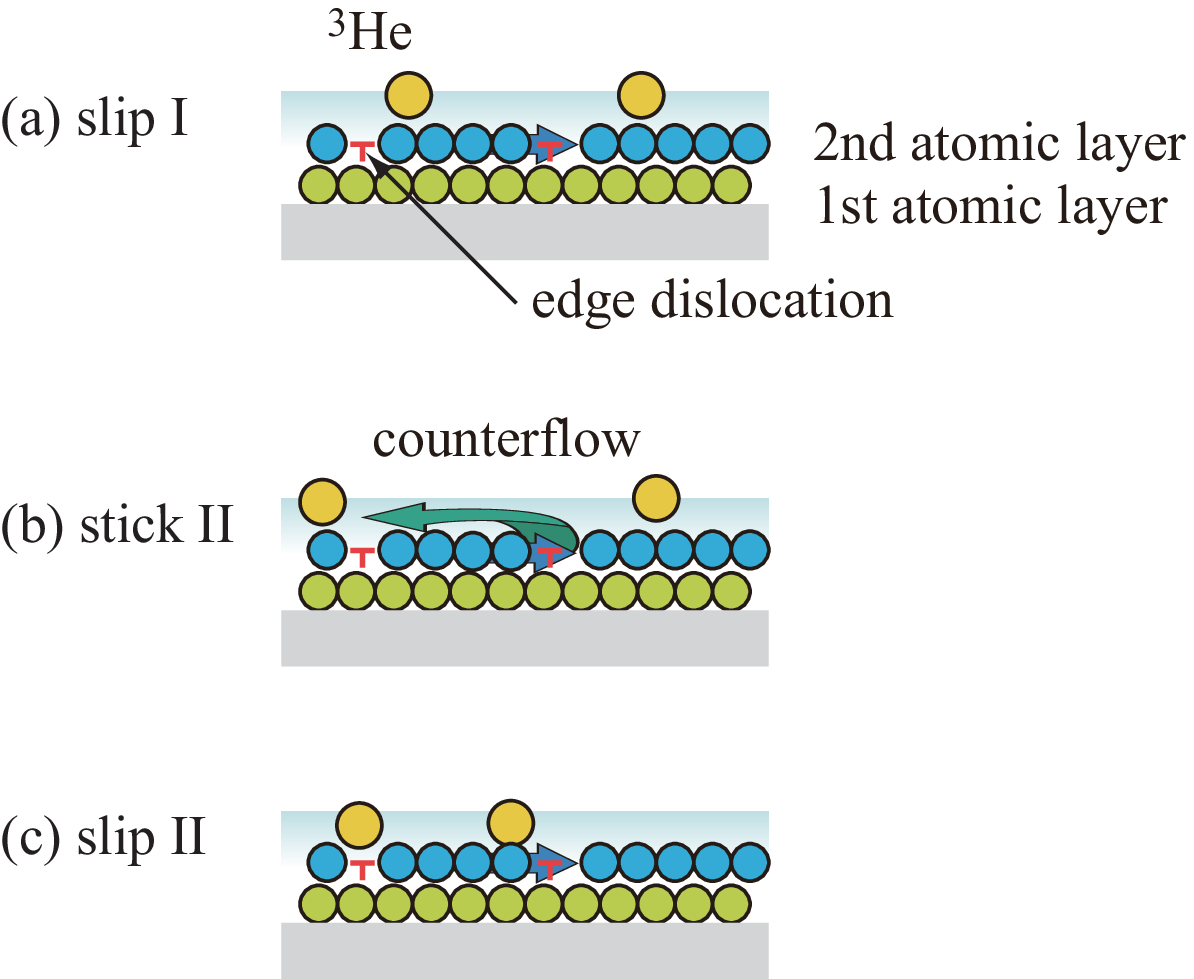}}
\caption{\label{fig:fig4} 
	Cartoon for $^{3}$He-$^{4}$He mixture films, (a) slip~I, 
	(b) stick~II and (c) slip~II. } 
\end{figure}

To clarify the $^{3}$He areal density dependence of $T_{S}$, $T_{D}$ and 
$T_{R}$, we carried out temperature sweep experiments with the amplitude 
of 0.25~nm for several $^{3}$He areal densities. 
Figure~3 shows the variation in resonance frequency. 
All data were taken during warming. When $^{3}$He of 0.05~atoms/nm$^{2}$ 
is added, the decoupling and sticking behaviors are drastically 
changed from the pure $^{4}$He film. 
$T_{S}$ is lowered down to 0.69~K from 0.74~K of the pure $^{4}$He film. 
In contrast, $T_{D}$ of 0.50~K does not change greatly. As the temperature 
decreases, the resonance frequency increases gradually below 0.4~K, and 
rises up at $T_{R}$ of 0.33~K. With further decreasing temperature down 
to 0.2~K, it decreases gradually again. Above $^{3}$He of 0.10~atoms/nm$^{2}$, 
the increase in frequency at $T_{R}$ becomes sharp. 
As the $^{3}$He areal density increases, $T_{R}$ increases gradually 
and $T_{D}$ does not change greatly. 

The inset shows a phase diagram of decoupling and sticking behaviors. 
This diagram is divided into four regions. At high temperature, the $^{4}$He 
solid layer sticks to the oscillating substrate (stick~I). 
As the temperature decreases, this layer undergoes decoupling below $T_{S}$ 
(slip~I), and sticks suddenly at $T_{D}$ (stick~II), regardless whether 
or not the film contains $^{3}$He. By adding $^{3}$He, the reentrant 
mass decoupling appears below $T_{R}$ (slip~II). 

We discuss a possible mechanism of the reentrant mass decoupling. 
It should be noted that $T_{D}$ and $T_{R}$ show up when the 
overlayer of these films becomes superfluid. For the pure $^{4}$He film, 
the sticking at $T_{D}$ can be explained by a mechanism in which the 
mass transport caused by the motion of edge dislocations is cancelled 
by the superfluid counterflow of the overlayer.\cite{Hosomi_2009}

In developing this scenario, we can explain the reentrant mass decoupling. 
A cartoon for $^{3}$He-$^{4}$He mixture films is shown in Fig.~4. 
Since $^{3}$He atoms are dissolved, or are spread over the fluid overlayer
 at high temperature, the $^{4}$He solid 
atomic layer shows the decoupling at $T_{S}$ and the sticking at $T_{D}$ 
as the same manner as pure $^{4}$He film (Figs.~4(a) and (b)). 
As above-mentioned, the sticking at $T_{D}$ means 
that the superfluid counterflow between edge dislocations cancels the 
mass transport. As the temperature decreases, $^{3}$He atoms start to 
adsorb on the edge dislocation at around $T_{3}$, and prevent the 
exchange between liquid and solid $^{4}$He atoms (Fig.~4(c)), 
i.e., the superfluid counterflow is ceased. Then, the $^{4}$He solid 
atomic layer undergoes decoupling again at $T_{R}$. 

\subsection{$^{4}$He areal density dependence}

In a different series of experiments from III.1., we carried out 
temperature sweep experiments for a fixed amount of $^{3}$He 
by changing the $^{4}$He areal density (Run~B). 
Figure~5 shows the variation in resonance frequency for several $^{4}$He 
areal densities with $^{3}$He of 0.20~atoms/nm$^{2}$. The oscillation 
amplitude was fixed at 0.18 and 0.018~nm during the temperature sweep. 

In the amplitude of 0.18~nm, all data were taken during warming. 
For $^{4}$He of 28.5~atoms/nm$^{2}$, the resonance frequency increases 
at $T_{S}$ of 0.64~K, which can be attributed to the mass decoupling 
of the $^{4}$He solid layer. For $^{4}$He of 29.0~atoms/nm$^{2}$, 
the mass decoupling is terminated abruptly at $T_{D}$ of 0.52~K and 
the reentrant mass decoupling occurs at $T_{R}$ of 0.39~K, as the same 
manner as Figs.~2~and~3. As the $^{4}$He areal density increases, 
the mass decoupling connected to $T_{S}$ disappears, while $T_{R}$ remains. 
As further increasing the $^{4}$He areal density, 
$T_{R}$ shifts to the lower temperature and vanishes at around $^{4}$He 
of 39.0~atoms/nm$^{2}$. 

In the amplitude of 0.018~nm, all data were taken during cooling. 
For $^{4}$He of 28.5~atoms/nm$^{2}$, it is difficult to definitely determine 
$T_{3}$, i.e., the sticking temperature of $^{3}$He atoms,  
while it is observed at $T_{3}$ of 0.40~K for $^{4}$He of 29.0~atoms/nm$^{2}$. 
As the $^{4}$He areal density increases, $T_{3}$ shifts to the 
lower temperature. Above $^{4}$He of 35.0~atoms/nm$^{2}$, it is difficult 
to determine $T_{3}$ again. Here, it should be noted that $T_{3}$ has the 
same $^{4}$He areal density dependence as $T_{R}$, although $T_{3}$ is 
observed in the limited range. 
On the other hand, the superfluid onset is also observed for 
these areal densities. As the $^{4}$He areal density increases, 
$T_{C}$ moves to the higher temperature and reaches 1.23~K 
at $^{4}$He of 39.0~atoms/nm$^{2}$.

\begin{figure}[t]
\centerline{\includegraphics[width=2.8in]{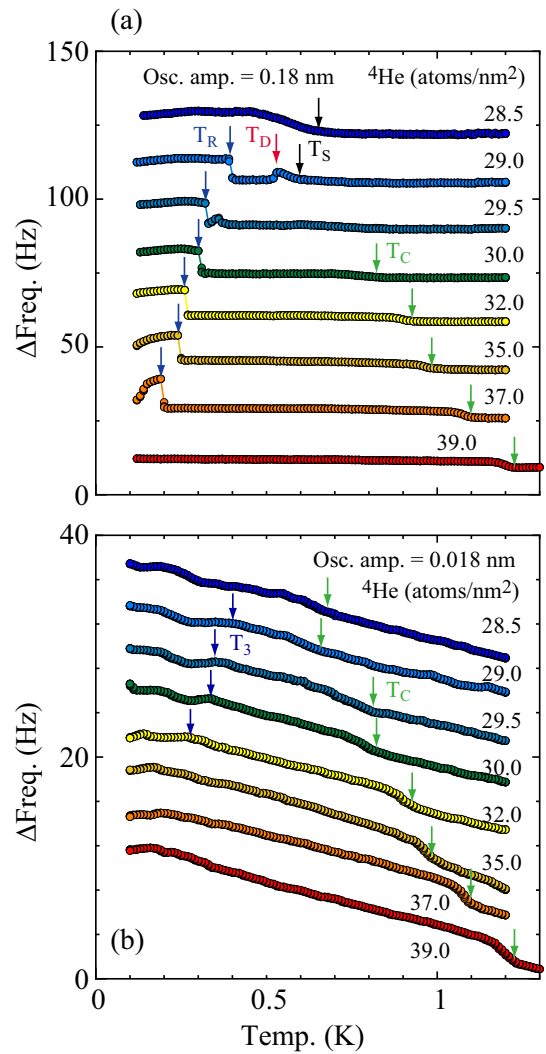}}
\caption{\label{fig:fig5} 
	Variations in the resonance frequency at the amplitudes of 
	(a) 0.18~nm and (b) 0.018~nm for various $^{4}$He areal 
	densities with $^{3}$He of 0.20~atoms/nm$^{2}$. 
	The data are shifted vertically. (Run~B)
	} 
\end{figure}

Figure 6 shows the phase diagram of the sticking and decoupling 
behaviors for $^{3}$He of 0.20~atoms/nm$^{2}$. $T_{S}$, $T_{D}$ 
and $T_{R}$ are obtained from the amplitudes of 0.18~nm, 
while $T_{C}$ and $T_{3}$ at the amplitudes of 0.018~nm. 
In contrast to the phase diagram of Fig.~3, both regions of 
Stick~I~and~II and of Slip~I~and~II connect continuously. 
As mentioned above, $T_{3}$ is close to $T_{R}$. This supports 
strongly the scenario mentioned in III.1., i.e., 
the adsorption of $^{3}$He atoms on the edge dislocation causes
the reentrant mass decoupling. 

Furthermore, the vanishment of $T_{R}$ at a high $^{4}$He areal 
density may be explained by the competition between the adsorption 
on the edge dislocation and on the free surface. For bulk $^{4}$He, 
it is well known that $^{3}$He atoms are bounded on the free surface 
at low temperature.\cite{Edwards_1978} The bound energy primarily 
comes from the difference in the zero-point energy between in bulk 
$^{4}$He liquid and on the free surface. 
Thus, we can propose the following scenario: 
In the case of an atomic-thin overlayer, $^{3}$He atoms are located 
on the $^{4}$He solid layer because of no advantage of the zero-point energy 
on the free surface. As the $^{4}$He areal density increases, i.e., 
the overlayer becomes thick, $^{3}$He atoms move to the free 
surface, and the adsorption of $^{3}$He atoms no longer occurs.

\begin{figure}
\centerline{\includegraphics[width=2.8in]{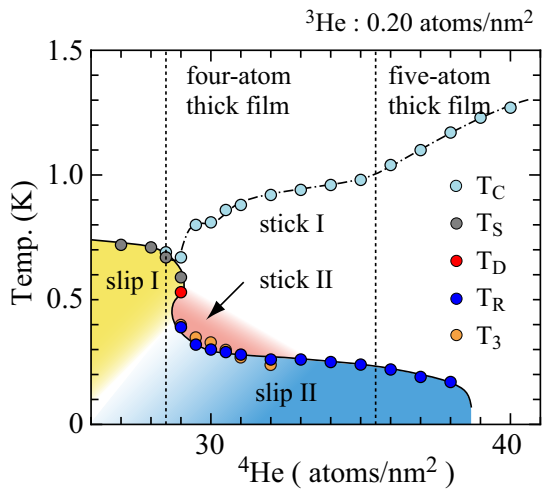}}
\caption{\label{fig:fig6} 
	Phase diagram of the sticking and decoupling behaviors 
	for $^{3}$He of 0.20~atoms/nm$^{2}$. $T_{S}$, $T_{D}$ and 
	$T_{R}$ are taken from the amplitudes of 0.18~nm, 
	$T_{C}$ and $T_{3}$ from the amplitudes of 0.018~nm. 
	} 
\end{figure}

\subsection{The model calculation for $^{3}$He adsorption}

We discuss whether the $^{3}$He areal density dependence of $T_{3}$ 
can be explained by a simple adsorption model. To build the model, 
we can refer the previous experiments for $^{3}$He-$^{4}$He mixture thin 
films.\cite{Dann_1998,Sprague_1995}  

Saunders and co-workers have carried out heat capacity experiments of $^{3}$He 
above 0.4~atoms/nm$^{2}$ in a $^{4}$He film of 33.5~atoms/nm$^{2}$ on Grafoil. 
They have reported that $^{3}$He atoms in a thin $^{4}$He film behave as 
the two-dimensional (2D) Fermi gas.\cite{Dann_1998} 
On the other hand, Hallock and co-workers have carried out NMR experiments 
for 0.1 monolayer of $^{3}$He in thin $^{4}$He films on Nuclepore.\cite{Sprague_1995}
They have reported that a part of $^{3}$He atoms are immobile below a critical 
$^{4}$He areal density. As the $^{4}$He areal density increases, $^{3}$He 
atoms experience a mobility edge. 

The present observations are quite similar to those of Hallock 
and co-workers' experiments, i.e., a small amount of $^{3}$He atoms 
are localized in thin $^{4}$He films, and this localization vanishes
at a certain $^{4}$He areal density. 
Sanders and co-workers concluded that $^{3}$He atoms in a thin $^{4}$He 
film are not adsorbed on Grafoil and are extended. 
We think, however, that there is a possibility that a small amount of $^{3}$He 
atoms are adsorbed because the heat capacity is independent of the areal 
density of the 2D Fermi gas. 

From these considerations, we consider the following model: $^{3}$He atoms 
in the overlayer behave as the 2D Fermi gas with the hydrodynamic effective 
mass $m_{3}^{*}$. In addition, there exits a surface binding state with 
the adsorption site density $N_{a}$ and the binding energy $\varepsilon_{a}$ 
measured from the bottom of the 2D Fermi gas. In this model, the adsorption 
density $n$ is obtained as
\begin{equation}
	n=N_{a}\frac{e^{-\beta(-\varepsilon_{a}-\mu)}}{1+e^{-\beta(-\varepsilon_{a}-\mu)}}
	+\frac{2}{(2\pi)^{2}}\int_0^{\infty}\frac{2\pi k\:dk}{e^{\beta(\varepsilon-\mu)}+1},
\end{equation}
where $\beta=1/k_{B}T$ is the inverse temperature, 
$\varepsilon = \hbar^{2}k^{2}/2m_{3}^{*}$ is the kinetic energy of the 
Fermi gas, and $\mu$ is the chemical potential which is determined from 
the $^{3}$He areal density. 

\begin{figure}
\centerline{\includegraphics[width=2.8in]{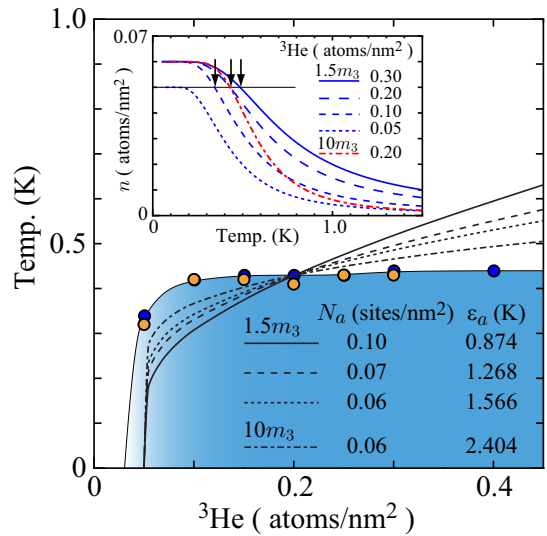}}
\caption{\label{fig:fig7} 
	Comparison between the model calculation and $T_{3}$. Open and solid 
	circles denote $T_{3}$ in Fig.~1 and $T_{R}$ in Fig.~3, respectively. 
	Again the $^{4}$He areal density is 29.3~atoms/nm$^{2}$. 
	The lines represent the temperature at which $n$ equals to 
	0.05~atoms/nm$^{2}$ with several different parameters. In calculation, 
	the parameters were chosen where the lines pass through 0.43~K at 
	0.20~atoms/nm$^{2}$. 
	} 
\end{figure}

Here, we may adopt $m_{3}^{*}/m_{3} \sim 1.5$ from heat capacity experiments 
for a $^{4}$He film of 33.5~atoms/nm$^{2}$,\cite{Dann_1998} although $m_{3}^{*}$ 
in thinner $^{4}$He films is still unknown. 
The inset of Fig.~7 shows a typical calculation of $n$ as a function of 
temperature for several $^{3}$He areal densities with $m_{3}^{*}/m_{3} = 1.5$, 
$N_{a} = 0.06$~sites/nm$^{2}$ and $\varepsilon_{a} = 1.268$~K. 
Here, the parameters were chosen where $n=0.05$~atoms/nm$^{2}$ 
at 0.43~K for $^{3}$He of 0.20~atoms/nm$^{2}$. 
For comparison, we plotted a curve of $m_{3}^{*}/m_{3} = 10$ at 0.20~atoms/nm$^{2}$. 
%
%

As seen in the inset, $n$ increases gradually from high temperature and 
becomes nearly equal to $N_{a}$ below a certain temperature. As the $^{3}$He 
areal density increases, $n$ shifts to the higher temperature. 
Although it is not clear that which value of $n$ corresponds to $T_{3}$, it
is assumed here that $n_{C} = 0.05$~atoms/nm$^{2}$ is $T_{3}$. 
We plotted the temperature where $n = 0.05$~atoms/nm$^{2}$ as a function of 
$^{3}$He areal density in Fig.~7. It was found that the calculated lines has 
a stronger $^{3}$He areal density dependence than that of the observations. 
This behavior does not depend strongly on the parameters of $N_{a}$, 
$\varepsilon_{a}$ and $n_{C}$. Thus, we may conclude that the simple 
adsorption model does not explain the areal density dependence of $T_{3}$. 

Here, we would like to make a comment on the model. As shown in Fig.~7, 
when we choose $m_{3}^{*}/m_{3} = 10$, $n$ varies rapidly in a small 
temperature range and the areal density dependence of $T_{3}$ becomes weaker.
i.e., when the number of density just above the surface binding state is 
large enough, $T_{3}$ does not depend strongly on $^{3}$He areal density. 
This may suggest that $^{3}$He atoms in a very thin overlayer are nearly 
localized on Grafoil. Although it was reported that $m_{3}^{*}$ is enhanced 
with decreasing $^{4}$He areal density for Nuclepore.\cite{Sprague_1995}, 
this is only speculation for Grafoil. Furhermore, when there exists an 
atractive interaction between the adsorption sites, $n$ varies more rapidly. 
These are for future study. 

Here, we make a comment on a thicker overlayer. $T_{3}$ was not 
observed clearly above $^{4}$He of 33.0~atoms/nm$^{3}$. It is, however, 
natural that $T_{3}$ is nearly equal to $T_{R}$. This means that 
$T_{3}$ tends to zero around $^{4}$He of 39.0~atoms/nm$^{3}$. 
As mentioned in III.2., $^{3}$He atoms are bounded on the free surface 
of bulk $^{4}$He.\cite{Edwards_1978} The binding energy $\varepsilon_{S}$ 
was obtained to be 2.22$\pm$0.03~K. This means that $\varepsilon_{a}$ 
in the simple adsorption model is smaller than $\varepsilon_{S}$ if the model 
explains the $^{3}$He adsorption. 

\section{Summary}

We report quartz crystal microbalance experiments using a 5~MHz AT-cut 
crystal for $^{3}$He-$^{4}$He mixture films on Grafoil. In the present 
experiments, the $^{3}$He areal density is at most up to 0.4 atoms/nm$^{2}$.
In a four-atom thick $^{4}$He film of 29.3~atoms/nm$^{2}$, we observed 
following behaviors: (a) For a small amplitude of 0.018~nm, a small drop 
in resonance frequency occurs at $T_{3}$.
(b) For a large amplitude of 0.25~nm, the mass decoupling at $T_{S}$ and 
sticking at $T_{D}$ was observed as the same manner as pure $^{4}$He 
films. In addition to $T_{S}$ and $T_{D}$, a reentrant mass decoupling 
occurs at $T_{R}$ close to $T_{3}$. Here, it was found that both of 
$T_{3}$ and $T_{R}$ do not depend strongly on the $^{3}$He areal density 
above 0.1~atoms/nm$^{2}$, and are $\sim$0.4~K. 

From our previous study for pure $^{4}$He films,
we have proposed the following scenario: the mass decoupling below $T_{S}$ 
results from the motion of edge dislocations between the first and second 
solid layers. 
The mass sticking at $T_{D}$ is caused by the cancellation of mass transport 
due to the superfluid counterflow of the overlayer.\cite{Hosomi_2009} 
As an extension of this scenario, the observed behaviors can be 
explained as follows. 
$^{3}$He atoms which are mobile at high temperature are localized on the 
edge dislocation at $T_{3}$. These $^{3}$He atoms prevent the exchange 
between liquid and solid $^{4}$He atoms, and the reentrant mass decoupling 
occurs by the cease of the superfluid counterflow. 

From experiments changing the $^{4}$He areal density for $^{3}$He of 
0.2~atoms/nm$^{2}$, it was found that $T_{R}$ decreases with increasing 
$^{4}$He areal density and vanishes above $^{4}$He of 29.0~atoms/nm$^{2}$. 
This behavior can be interpreted by the competition of the adsorption between 
on the edge dislocation and on the free surface. 

The above explanation about the reentrant mass decoupling below $T_{R}$ 
naturally leads to the model that $^{3}$He atoms adsorb on the $^{4}$He 
solid layer. However, this model cannot explain a weak $^{3}$He 
areal density dependence of $T_{R}$ using the known hydrodynamic effective 
mass of $^{3}$He in the overlayer. This is for future study. 

\begin{acknowledgments}

One of the authors (TM) wishes to express his thanks for the financial 
support of Yamaguchi Educational and Scholarship Foundation.

\end{acknowledgments}


\end{document}